\documentclass[a4paper,10pt]{article}
\usepackage{enumerate}
\usepackage{color}
\usepackage[utf8]{inputenc} 
\usepackage[english]{babel}
\usepackage[T1]{fontenc}
\usepackage{graphicx}
\usepackage{amsfonts,amssymb,amsmath,latexsym,amsthm}
\usepackage{textcomp}
\usepackage[pdftex]{hyperref}
\usepackage{geometry}
\title{Evolution and statistical analysis of random wave fields within the Benjamin-Ono equation}
\author{Marcelo V. Flamarion$^{1}$ and Efim Pelinovsky$^{2,3}$}
\date{}

\begin{document}
\maketitle
\begin{center}
{\footnotesize $^1$Unidade Acad{\^ e}mica do Cabo de Santo Agostinho, \\
UFRPE/Rural Federal University of Pernambuco, BR 101 Sul, Cabo de Santo Agostinho-PE, Brazil,  54503-900 \\
marcelo.flamarion@ufrpe.br }

\vspace{0.3cm}
{\footnotesize $^{2}$Institute of Applied Physics, 46 Uljanov Str., Nizhny Novgorod 603155, Russia. \\
 $^{3}$National Research University--Higher School of Economics, Moscow, Russia. }


\end{center}


\begin{abstract} 
This study investigates the numerical evolution of an initially internal random wave field characterized by a Gaussian spectrum shape using the Benjamin-Ono (BO) equation. The research focuses on analyzing various properties associated with the BO random wave field, including the transition to a steady state of the spectra, statistical moments, and the distribution functions of wave amplitudes. Numerical simulations are conducted across different Ursell parameters, revealing intriguing findings. Notably, it is observed that the spectra of the wave field converges to a stationary state in a statistical sense, while exhibiting statistical characteristics that deviate from a Gaussian distribution. Moreover, as the Ursell parameter increases, the positive skewness of the wave field intensifies, and the kurtosis increases. The investigation also involves the computation of the probability of rogue wave formation, revealing deviations from the Rayleigh distribution. Notably, the study uncovers distinct types of rogue waves, specifically referred to as "two sisters" and "three sisters" phenomena.

	\end{abstract}

\section{Introduction}
Ocean internal waves are fascinating phenomena characterized by underwater oscillations with amplitudes ranging from 50 to 100 meters, and sometimes even more \cite{Mignerey:2003, Duda:2004, Apel:2007}. These waves are generated when there is a disturbance at the interface between water layers of varying densities. This disturbance can propagate over vast distances, causing the layers to oscillate in relation to their initial equilibrium state. One of the main causes of internal waves is the deterministic mechanism triggered by tidal flows interacting with bathymetric features like seamounts or continental shelves, other mechanisms may include instability of ocean currents in zones with strong shear flow (such as an ocean gulf), or directly by wind stress \cite{Grimshaw:2010}. This phenomenon is commonly referred to as an internal tide. As the tides flow over these underwater obstacles, they create disturbances that propagate as internal waves throughout the ocean. In addition to deterministic mechanisms, internal waves can also be excited by random events such as tsunamis or severe ocean storms. These powerful and unpredictable phenomena have the ability to generate significant disturbances in the ocean, leading to the formation of large-scale internal waves.

Extensive research has been conducted to investigate weakly nonlinear models that depict the behavior of internal waves. Notably, the Korteweg-de Vries (KdV) equation and the Gardner equation have received significant attention in this regard \cite{Grimshaw:2010a, Grimshaw:2010, Bokaeeyan:2019}. The KdV and Gardner equation are applicable to shallow water scenarios, while the Intermediate Long Wave (ILW) equation is suitable for fluids of finite depth \cite{Choi:1996, Matsuno:1993a, Matsuno:1993b}. Additionally, the Benjamin-Ono (BO) equation pertains to deep water dynamics  \cite{Benjamin:1967, Davis:1967, Ono:1975, Ko:1978, Choi:1996}. These well-established models exhibit intriguing features, such as the existence of periodic and solitary wave solutions that endure over time. However, it is crucial to acknowledge that these model equations have inherent limitations that constrain their applicability to more generalized problems.
\begin{figure}[h!]
	\centering	
	\includegraphics[scale =1]{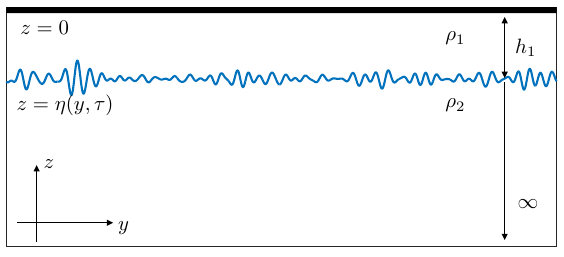}
	\caption{Sketch of the BO model.}
	\label{Fig1}
\end{figure}

The widely recognized Benjamin-Ono (BO) equation
\begin{equation}\label{BO0}
\eta_{\tau} +c_{0}\eta_y-\frac{3c_0}{2h_1}\eta\eta_{y}+\frac{c_0h_1}{2\rho_r}\mathcal{H}[\eta_{yy}]=0,
\end{equation}
is frequently employed to examine the behavior of a perturbed interface between two inviscid fluids. This interface is characterized by a flat rigid lid and infinite depth. In this equation, various parameters and functions play important roles. The thickness of the upper fluid layer with density $\rho_1$ is represented by $h_1$, while $\rho_2$ denotes the density of the lower fluid. The ratio of densities between the lighter upper layer and the heavier lower layer is denoted as $\rho_r=\rho_1/\rho_2<1$. The linear speed $c_0$ is 
\begin{equation}\label{speed}
c_0^2=gh_{1}\Big(\frac{1}{\rho_r}-1\Big),
\end{equation}
where $g$ is the acceleration of gravity. More details of the geometry of the problem is depicted in Figure \ref{Fig1}. The elevation of the interface in the position $y$ and time $\tau$ is denoted by $\eta(y,\tau)$ and $\mathcal{H}$ denotes the Hilbert transform defined as
\begin{equation}\label{Hilbert}
\mathcal{H}[\eta(y,\tau)]=\frac{1}{\pi}\int_{-\infty}^{+\infty}\frac{\eta(z,\tau)}{z-y}dz.
\end{equation}

The BO equation is used to model several problems such as ship wakes, flow of water over rocks, the formation of storms in the ocean and even atmospheric problems  such as atmospheric flows encountering obstacles \cite{Baines, Johnson}. These problems are typically approached as deterministic, where the initial state of the perturbed interface is precisely known, and the evolution of the phenomenon over time can be computed using the BO equations. However, in many cases, the scarcity of data for initializing deterministic models presents a challenge. In such instances, assuming random initial states becomes preferable, allowing for the study and formulation of statistically reliable predictions \cite{Didenkulova:2019, Dutykh:2014, Pelinovsky:2006, Pelinovsky:2016, Viotti:2013}. 
Analogously, the medium in which a wave propagates can also be considered random \cite{Alfaro-Vigo:2004, Fouque:2004, Fouque:2005, Solna}.

In the context of surface waves, the study conducted by Pelinovsky and Sergeeva \cite{Pelinovsky:2006} focused on numerically examining the evolution of an initially random wave field with a Gaussian spectrum shape using the KdV equation. Their findings revealed that the irregular wave field, resulting from the wave evolution within the KdV model, does not adhere to Gaussian statistics. Instead, its statistical properties are influenced by the Ursell parameter, which represents the ratio of nonlinear effects to dispersion. Consequently, the wave field becomes asymmetric, exhibiting sharper crests, leading to a positive third moment. Importantly, the study demonstrates the existence of a steady state for statistical characteristics, including skewness, kurtosis, distribution functions, and spectral density. Through computations, it was observed that both statistical moments and distribution functions evolve until they reach a certain bound level. A similar effect is observed in the evolution of a random wave spectrum. In a related study, Didenkulova et al. \cite{Didenkulova:2019} conducted direct numerical simulations of nonlinear wave evolution within the framework of the KdV equation for cases involving bimodal wave spectra models. They investigated the coexistence effect of an additional wave system on the evolution of wave statistical characteristics, spectral shapes, and the resulting equilibrium state. Furthermore, the study demonstrated that the presence of a low-frequency spectral component leads to more asymmetric waves with more extreme statistical properties..

The objective of this study is to investigate the behaviour of the evolution of a random internal wave field through the BO equation. As our model is deterministic, we conduct numerous simulations  with randomly generated initial conditions with identical statistical characteristics. For the mechanism of turbulence on  internal waves the readers are referred to \cite{Lvov1, Lvov2, Lvov3, Lvov4}. Our approach is related with a case of integrable turbulence of internal waves because the BO equation which is integrable model. Through a series of numerical simulations we investigate spectrum evolution and stabilization, the third and the fourth statistical moments of the random wave field, and the distribution function for the crest amplitudes. In addition, we pay special attention to the formation of "rogue internal waves" finding the so-called ``three sister" and other extensions.

For reference, this article is organized as follows: The O equation is presented in Section 2. In Section 3, we describe the numerical methods and the results on Section 4. Then, we present the final considerations in Section 5.

\section{The Benjamin-Ono equation}
Our study aims to investigate the underlying cause of the dominant dynamic, focusing on the nonlinearity of dispersion. To achieve this, it is convenient to utilize dimensionless variables by introducing the BO equation. Equation (\ref{BO0}) is typically initialized with 
\begin{equation}\label{IW}
\eta(y,0)=A_{s}F(Kk_{0}y)\sin(k_{0}y).
\end{equation}
Here, $A_s$ represents a standard wave amplitude, $F$ denotes the wave envelope with spectrum width $K$, and $k_0$ corresponds to the carrier wave number. Thus, we introduce dimensionless variables
\begin{equation}\label{dimensionless}
x = k_{0}(y-c_{0}t), \;\ t = \frac{3c_0k_0A_s}{2h_1}\tau \mbox{ and } \eta(y,\tau) = -\frac{u(x,t)}{A_s}.
\end{equation}

Substituting the dimensionless variables (\ref{dimensionless}) into equation (\ref{BO0}), we derive the dimensionless BO equation
\begin{equation}\label{BO}
u_{t} +uu_{x}+\frac{1}{U_{r}}\mathcal{H}[u_{xx}]=0,
\end{equation}
where $U_r$ represents the Ursell parameter, given by $U_r=A_{s}\rho_{r}/h_{1}k_{0}$. This parameter governs the nonlinearity of the BO equation, with higher values of $U_r$ indicating predominantly nonlinear dynamics, while lower values of $U_r$ suggesting predominantly linear dynamics. Additionally, the Ursell parameter depends on the thickness of upper layer, hence variations on the thickness of the upper layer may change the wave regime significantly. 

To incorporate randomness into the problem, we introduce a zero-mean random state modeled as a Fourier series with $M$ harmonics. The initial condition is given by
\begin{equation}\label{initial}
u(x,0) = \sum_{i=1}^{M}\sqrt{2S(k_i)\Delta k}\cos(k_{i}x+\varphi_{i}),
\end{equation}
Here, $S(k)$ represents the initial power spectrum, $k_i = i\Delta k$ with $\Delta k$ being the sampling wave number, and $\varphi_i$ denotes a random variable uniformly distributed in the interval $(0,2\pi)$. The length of the initial realization is $L = 2\pi/\Delta k$. We assume that the initial power spectrum follows a Gaussian distribution:
\begin{equation}\label{Gaussian}
S(k) = Q\exp\Big(-\frac{1}{2}\frac{(k-k_0)^2}{2K^2}\Big), \;\ k>0.
\end{equation}

The wave characteristics are determined by several parameters: the dimensionless peak wavenumber $k_0=1$, the spectral width denoted as $K=0.18$, and the relative energy indicated by the values of $Q$. All scenarios share the same total energy of the waves, which is defined by the variance $\sigma^2 = 0.25$
\begin{equation}\label{sigma}
\sigma^{2}=\frac{1}{L}\int_{0}^{L}u^{2}(x,t)dx.
\end{equation}
Within the integrable BO equation, the variance remains constant throughout the motion. For the spectral domain, we choose a dimension of $256$ harmonics, allowing for a spectrum decay in the higher $k$ region. These parameter choices align with the ones utilized by Pelinovsky and Sergeeva \cite{Pelinovsky:2006} and Didenkulova et al. \cite{Didenkulova:2019}. By employing these specific parameters, we can effectively compare the results presented in the subsequent sections with their findings.

\section{Numerical methods}
The numerical solution of equation (\ref{BO}) is obtained using a Fourier pseudospectral method with an integrating factor. The equation is solved in a periodic computational domain of $[0, 2\pi/\Delta k]$, where $\Delta k = 0.023$, and a uniform grid containing $N=2^{12}$ points is employed. This grid ensures accurate approximation of spatial derivatives \cite{Trefethen:2000}.  For the time evolution of the equation, we utilize the classical fourth-order Runge-Kutta method with discrete time steps of size $\Delta t=0.005$. For a more detailed resolution and comprehensive understanding of a similar numerical method, readers are referred to the work of Flamarion et al. \cite{Marcelo-Paul-Andre}. Furthermore, we use the fact    that the BO equation (\ref{BO}) conserves both the total mass $(m(t))$ and the momentum $(p(t))$ to verify the accuracy of the chosen numerical method. The total mass is defined as
\begin{equation}\label{mass}
\frac{dm}{dt}=0, \quad \text{where } m(t)=\frac{1}{L}\int_{0}^{L}u(x,t)dx,
\end{equation}
and the momentum is defined as
\begin{equation}\label{momentum}
\frac{dp}{dt}=0, \quad \text{where } p(t)=\frac{1}{L}\int_{0}^{L}u^{2}(x,t)dx.
\end{equation}
Numerical simulations are controlled by retaining of the first and second moments with precision of machine  and $10^{-9}$ respectively. Figure \ref{mass} illustrates the conservation of mass and momentum over time for the same simulations shown in Figure \ref{Mesh}. As observed, both quantities remain conserved throughout the simulations. Notice that for the mass conservation we have machine precision.
\begin{figure}[h!]
	\centering
	\includegraphics[scale =1.1]{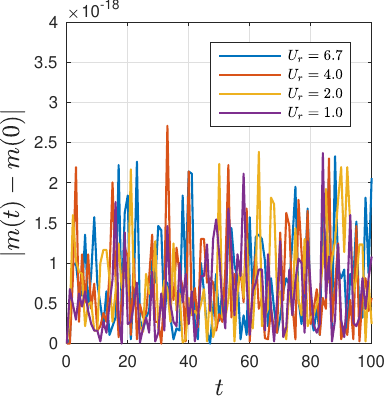}	
	\includegraphics[scale =1.1]{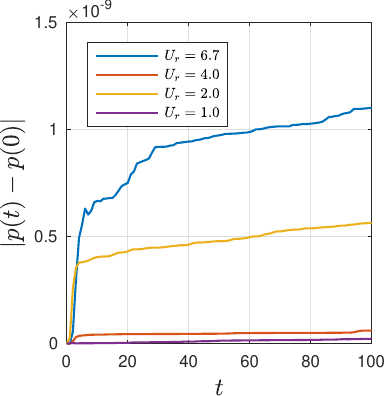}	
	\caption{ The conserved quantities of BO equation of the simulations shown in Figure \ref{Mesh}. On the left the mass conservation and on the right the momentum.
}
	\label{mass}
\end{figure}

\section{Results}

\subsection{Wave field}
\begin{figure}[h!]
	\centering
	\includegraphics[scale =1.1]{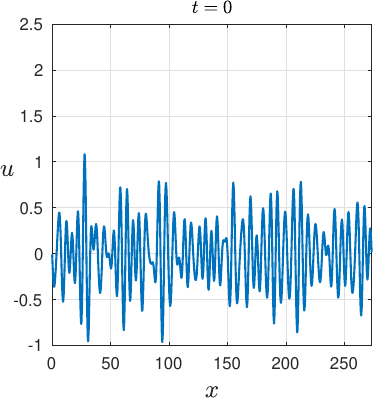}	
	\includegraphics[scale =1.1]{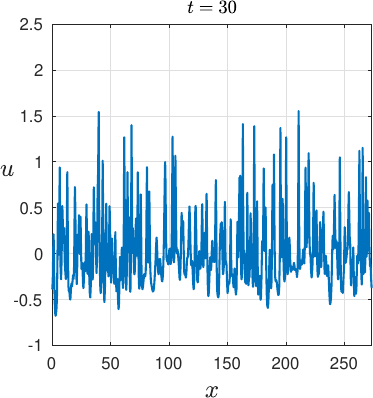}	
	\includegraphics[scale =1.1]{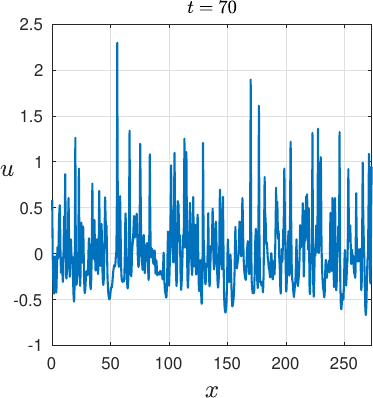}	
	\includegraphics[scale =1.1]{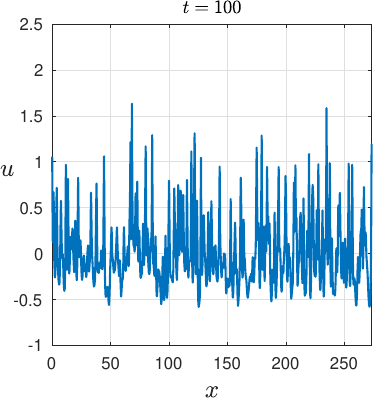}	
	\caption{ The evolution of the wave field at different times with $U_r=6.7$.}
	\label{Wavefield}
\end{figure}

The evolution of the wave record is depicted in Figure \ref{Wavefield} at various time instances. Initially, the wave field exhibits a relatively narrow spectrum, resulting in smooth group structures of the waves. However, as time progresses, the wave profile becomes asymmetric, transitioning from smooth crests and troughs to sharp ones. This signifies an increase in the skewness of the wave field beyond its initial value, indicating a departure from a symmetric distribution, which will be discussed later. Moreover, the chosen parameter $U_{r}=6.7$ corresponds to a significant nonlinearity, leading to the emergence of large-amplitude waves during the interaction.

\begin{figure}[h!]
	\centering
	\includegraphics[scale =1.1]{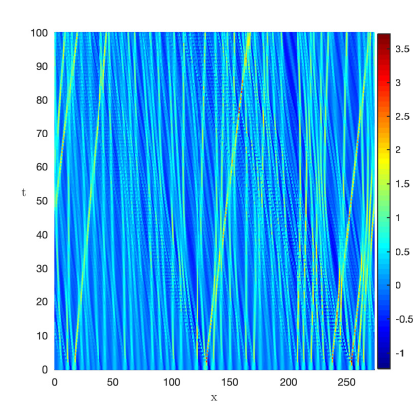}	
	\includegraphics[scale =1.1]{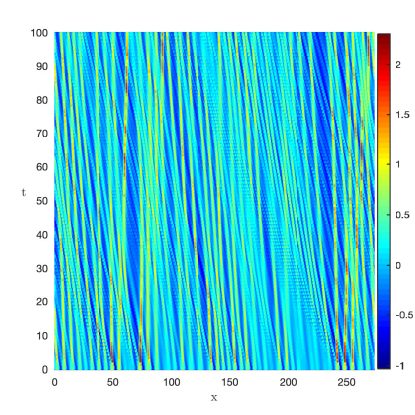}	
	\includegraphics[scale =1.1]{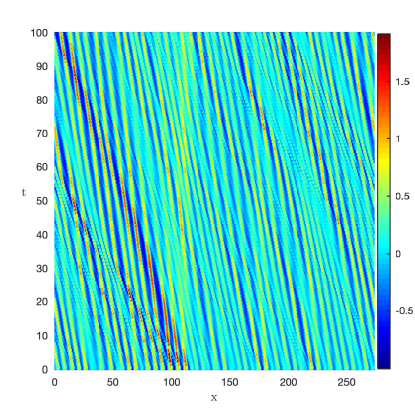}	
	\includegraphics[scale =1.1]{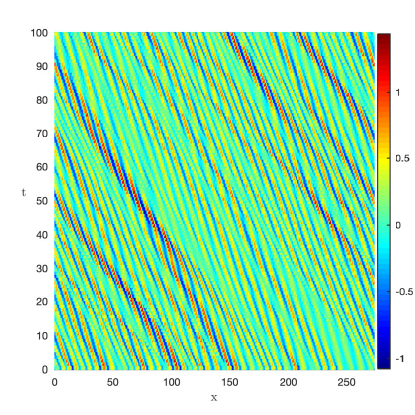}	
	\caption{ The evolution of the wave field  for different values of the parameter $U_r$. From top to bottom and left to right  $U_r=6.7$, $U_r=4.0$, $U_r=2.0$ and $U_r=1.0$.}
	\label{Mesh}
\end{figure}

\begin{figure}[h!]
	\centering
	\includegraphics[scale =1.1]{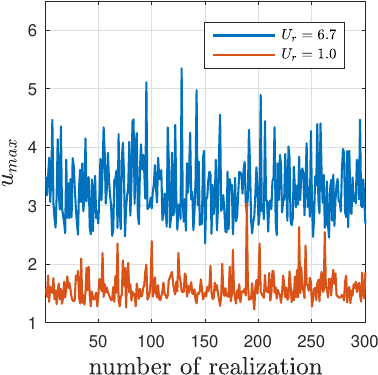}	
	\includegraphics[scale =1.1]{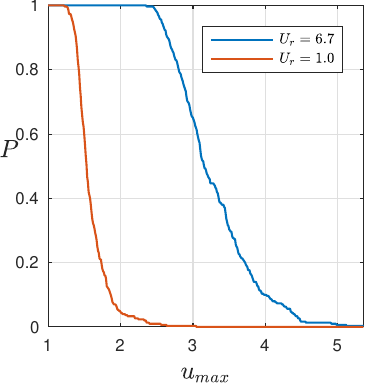}	
	\caption{ Left: Maximum of wave amplitudes over time.  Right: distribution of maximum crest amplitudes over $300$ realizations for different values of $U_r$.}
	\label{crests}
\end{figure}

To comprehend the role of the Ursell parameter in the dynamics, we analyze the trajectory patterns displayed in Figure \ref{Mesh}. On the time-space plane, we observe that for higher values of $U_r$, the wave propagation results in the formation of prominent peak amplitudes, as indicated in Figure \ref{Mesh}. The  the distribution of crest amplitudes is presented in Figure \ref{crests} shows that nonlinearity is essential for the existence of waves with large amplitudes, in particular in the formation of rogue waves, which will be discussed below.

\begin{figure}[h!]
	\centering
	\includegraphics[scale =1.1]{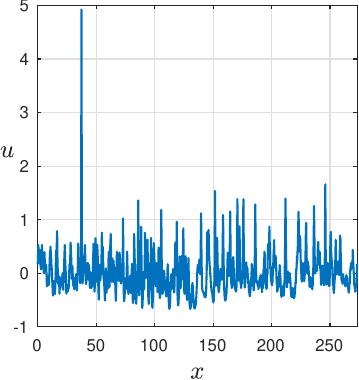}	
	\includegraphics[scale =1.1]{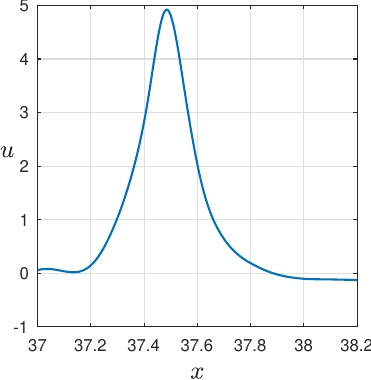}	
	\includegraphics[scale =1.1]{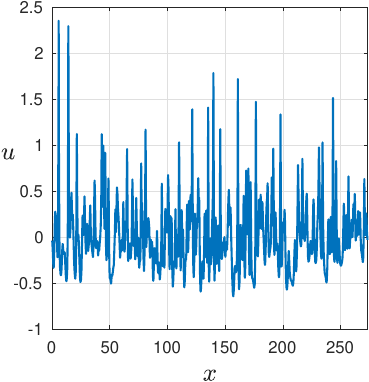}	
	\includegraphics[scale =1.1]{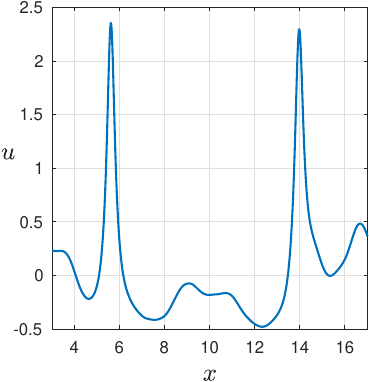}	
	\includegraphics[scale =1.1]{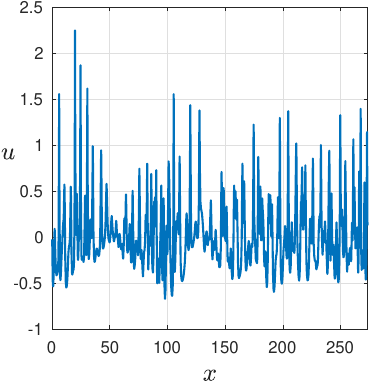}	
	\includegraphics[scale =1.1]{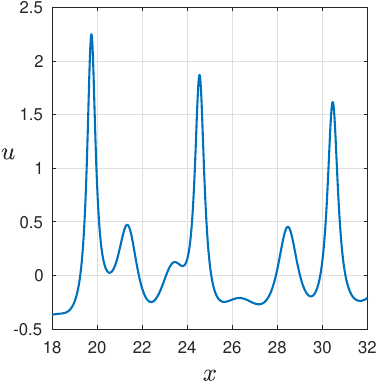}	
	\caption{Left: The occurrence of a freak wave, ``two sisters" and ``three sisters" at simulation number $142$, time $t=99$, number $4$, time $t=99$ and number $84$ and time $t=20$ respectively for $U_{r}=6.7$. Right: Zoom of the freak wave.}
	\label{Freak}
\end{figure}

In recent years, there has been a growing fascination with rogue waves, an intriguing type of nonlinear wave also known as freak waves. These waves have captured the attention of researchers from various scientific disciplines due to their unique characteristics. Initially observed in the deep ocean, rogue waves exhibit abnormal behavior, with amplitudes two to three times higher than the surrounding waves. What makes them particularly captivating is their sudden formation, seemingly appearing out of nowhere.

Rogue waves have become the subject of extensive investigation in fields such as oceanography \cite{Dudley:2019, Pelinovsky:Book}, optical fibers \cite{Akhmediev:2009, Solli:2009, Yeom:2007} , Bose-Einstein condensates \cite{Manikandan:2016, Tan:2022, Bai:2022}, financial markets \cite{Zhen-Ya:2010}, and related areas. Mathematically, the criterion for the occurrence of freak waves can be described by the equation \cite{Pelinovsky:Book}
\begin{equation}\label{freak}
A_{fr} > 2H_{s},
\end{equation}
where $A_{fr}$ represents the amplitude of the freak wave, and $H_s$ represents the amplitude of the "significant" wave field. The value of $H_s$ is determined by averaging one-third of the largest waves observed in the given context, such as in the field of oceanology.

Figure \ref{Freak} (left) displays different simulated wave fields at different time records, where prominent peaks significantly surpass the surrounding waves, representing various types of freak waves. Zooming in on the left panels, Figure \ref{Freak} (right) provides a closer look at their distinctive appearance. For instance, the top-right image showcases a freak wave of substantial amplitude known as the "two sisters" and "three sisters." It is worth mentioning that series of big waves with more than four peaks were also observed. The occurrence of freak waves is strongly influenced by the parameter $U_r$, with a higher frequency expected in wave fields characterized by stronger nonlinearity and smaller values of $U_r$.

By studying and understanding the characteristics and behavior of these rogue waves, researchers aim to shed light on their formation mechanisms and develop methods to predict and mitigate their potentially hazardous effects.

\subsection{Spectra}
\begin{figure}[h!]
	\centering
	\includegraphics[scale =1.1]{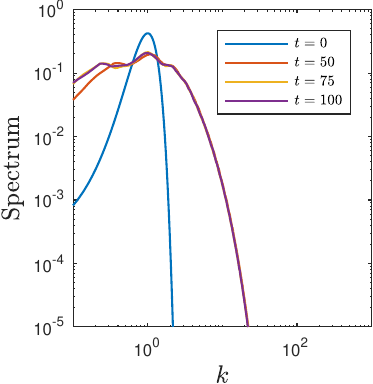}	
	\includegraphics[scale =1.1]{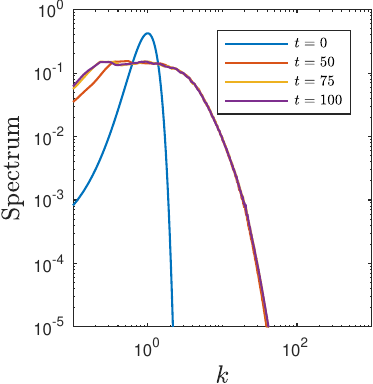}	
	\includegraphics[scale =1.1]{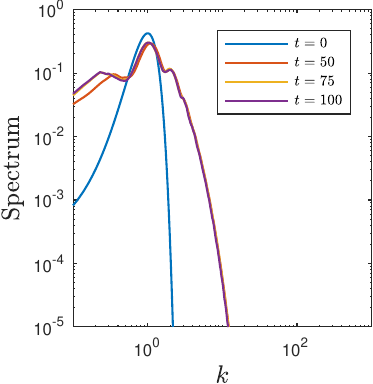}	
	\includegraphics[scale =1.1]{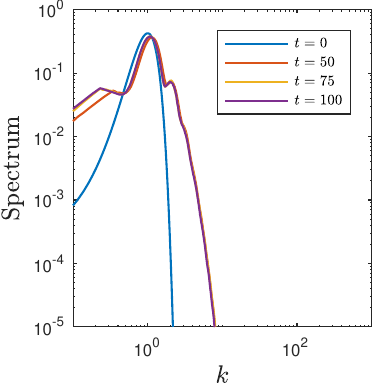}	
	\includegraphics[scale =1.1]{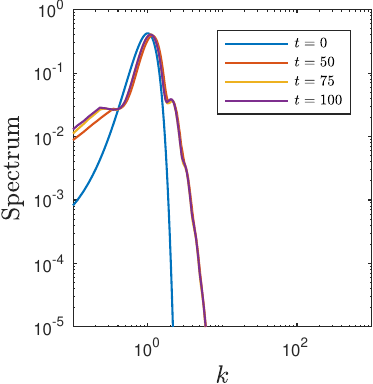}	
	\includegraphics[scale =1.1]{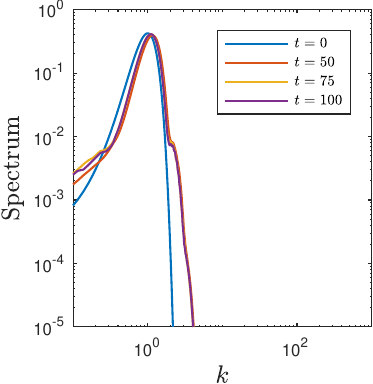}	
	\caption{Averaged evolution of spectra of the wave fields at different times. From top to bottom and left to right  $U_r=6.7$, $U_r=4.0$, $U_r=2.0$, $U_r=1.0$, $U_r=0.5$ and $U_r=0.1$.}
	\label{spectra}
\end{figure}

The evolution of the spectrum is studied across various $U_r$ values, ranging from $0.1$ (representing nearly linear progression) to $6.7$ (indicating a highly nonlinear wave behavior). To ensure reliable statistical outcomes, $300$ realizations are averaged over different time periods. As anticipated, the presence of nonlinearity triggers a transformation in the spectrum, leading to its widening and eventual convergence into a stable state (refer to Figure \ref{spectra}). The nature of this stable state, determined by the Ursell parameter, displays an asymmetrical shape with a noticeable shift of energy towards lower frequencies, commonly known as the spectrum downshift effect. In cases of larger Ursell parameter values (as illustrated in (refer to Figure \ref{spectra}) (top)), the spectral density becomes more evenly distributed across smaller wave numbers ($k$-values). The prominence of spectrum flatness intensifies under strong nonlinearity ($U_r = 6.7$), indicating a higher energy level in the wave field and an increased significance of nonlinear effects. Consequently, a broader range of frequencies is necessary to accurately represent the wave spectrum. This inclination towards spectrum flatness aligns with the concept of statistical equilibrium in the absence of external inputs or outputs. These findings corroborate and strengthen the outcomes reported by Pelinovsky and Sergeeva \cite{Pelinovsky:2006} within the framework of the KdV equation.

\subsection{Statistical moments}
To gain a deeper understanding of the wave fields interactions described in the previous subsection, we focus on examining four specific integrals, corresponding to four statistical moments
\begin{equation}\label{moments}
\mu_{n}(t)=\frac{1}{L}\int_{0}^{L}u^{n}(x,t)dx, \mbox{ where $n=1,2,3,4.$}
\end{equation}
More precisely we focus on two statistical quantities that characterize the wave spectrum. The kurtosis excess ($\kappa$) and the skewness ($\varsigma$) defined as
\begin{equation}\label{kurtskew}
\kappa(t) = \frac{\mu_{4}}{\mu_{2}^{2}}-3 \mbox{ and } \varsigma(t) = \frac{\mu_{3}}{\mu_{2}^{3/2}}.
\end{equation}

The kurtosis is a measure that indicates the tail heaviness of the spectrum. In simpler terms, it quantifies the degree of peakedness in the distribution and characterizes the influence of large waves on the overall distribution. A positive kurtosis implies a substantial contribution from large waves. Skewness, on the other hand, measures the asymmetry of the spectrum relative to the mean. Specifically, it represents the statistical measure of vertical asymmetry in the wave field, with its sign indicating the ratio of crests to troughs. A positive skewness signifies that crests are larger than troughs.

\begin{figure}[h!]
	\centering
	\includegraphics[scale =1.1]{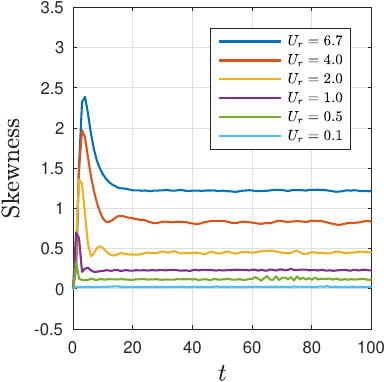}	
		\includegraphics[scale =1.1]{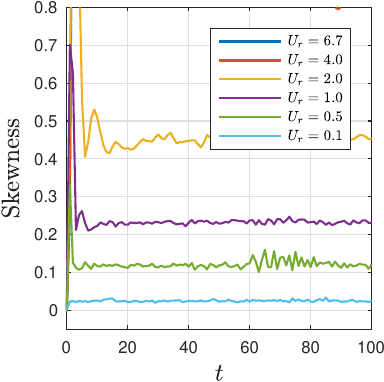}	
	\includegraphics[scale =1.1]{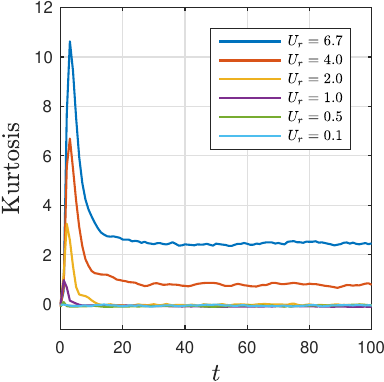}
		\includegraphics[scale =1.1]{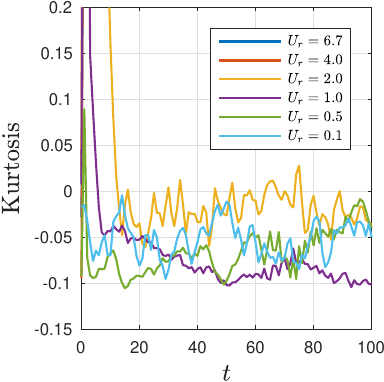}
	\caption{ Temporal evolution of the skewness and kurtosis for different values of the parameter $U_r$ averaged over 300 realizations. On the right, the zooming of the figures on the left.}
	\label{Skewness_Comp_Mean}
\end{figure}

The behavior of statistical moments provides insights into the presence of a stationary state and its evolution over time. During a transition period of approximately 10 to 20 nonlinear time units, both moments of the wave field converge towards nearly constant values (see Figure \ref{Skewness_Comp_Mean}). Regardless of the conditions, the skewness of the wave field remains positive, indicating that positive waves (crests) have larger amplitudes compared to negative waves (troughs). Furthermore, the asymptotic value of skewness increases as the Ursell parameter rises. 

When the Ursell number is $2$, the kurtosis shows oscillatory patterns centered around zero. This scenario resembles the findings reported by Onorato et al. \cite{Onorato} and Tanaka \cite{Tanaka} in the context of deep water. In their studies, it was demonstrated that in deep water, the calculated kurtosis values exhibit oscillations around zero. For highly nonlinear random wave processes, some values surpass zero, indicating an increased likelihood of encountering large-amplitude waves, including freak waves. On the other hand, for smaller values, the kurtosis becomes negative, suggesting a lower probability of encountering such extreme events compared to what is expected for Gaussian processes. Conversely, under strong nonlinearity, the asymptotic value of kurtosis exceeds zero, indicating a higher probability of encountering large waves, with significant contributions from smaller waves within the wave field.


\subsection{Crest distribution}
\begin{figure}[h!]
	\centering
	\includegraphics[scale =1.1]{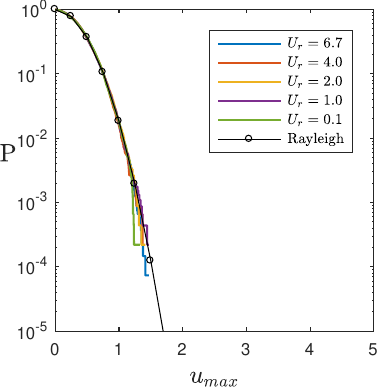}	
	\includegraphics[scale =1.1]{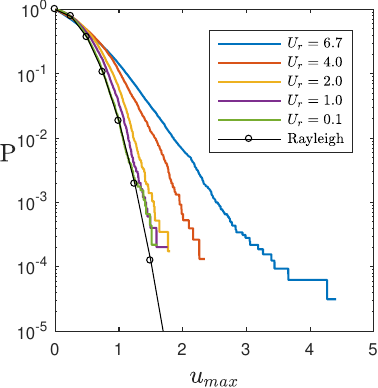}
	\caption{ The exceedance probability distributions for wave crests  at the initial moment ($t=0$) and the asymtotic distribution ($t=100$) (right) for different Ursell numbers over $300$ realizations. The black solid line with circles corresponds to the Rayleigh distribution of the narrow-band Gaussian process.}
	\label{Ray}
\end{figure}

In this section, we compare the exceedance probability distributions of wave crests for different $U_r$ values with the Rayleigh distribution of amplitudes for a narrow-band Gaussian process
\begin{equation}\label{RayII}
P(u_{max}) = \exp\Big(- \frac{u_{max}^{2}}{2\sigma^{2}}\Big).
\end{equation}
The instant distributions of wave crests, denoted as $u_{max}$ (representing local maxima of $u$ between consecutive zero-crossing points), are computed for a dataset comprising approximately 150,000 waves. Initially, the wave phases are randomly distributed, and the crest distributions, denoted as $P(u_{max})$, exhibit a reasonable agreement with the Rayleigh law for narrow-banded waves (refer to Figure \ref{Ray} (left)).

The probability distributions evolve over time. In the quasi-equilibrium stage (Figure \ref{Ray} (right)), the wave crest distributions are compared with the theoretical predictions provided by Equation (\ref{RayII}). Across all five cases, the distributions exhibit qualitatively similar behavior in relation to the reference Rayleigh curve for small amplitude waves. The asymptotic distribution surpasses the Rayleigh distribution, indicating an increased probability of encountering higher wave crests. Qualitatively, the shape of the amplitude distribution function aligns with the behavior of skewness and kurtosis depicted in Figure \ref{Skewness_Comp_Mean}. The positive waves exhibit larger amplitudes than the negative waves, as indicated by the skewness, while the kurtosis highlights the significant contribution of small waves to the overall distribution.

\section{Conclusion}
In this work, we have explored the statistical properties of a random internal wave field with initially identical statistical properties. Our findings under strong nonlinearity have revealed the existence of different types of freak waves, including the intriguing three sisters phenomenon. We have observed that the spectral evolution of the wave field reaches a steady state across all Ursell parameters, but with increased nonlinearity, the spectral density becomes more uniformly distributed. This outcome aligns with previous research conducted by Pelinovsky and Sergeeva \cite{Pelinovsky:2006}. Furthermore, we have computed the kurtosis and skewness of the wave field. The positive skewness values indicate the asymmetry of the wave field, with sharper crests. As for the kurtosis, strong nonlinearity (large Ursell parameter values) demonstrates a significant contribution from larger waves in the wave dynamics. In contrast, strong dispersion (small Ursell parameter values) leads to kurtosis oscillating around zero or becoming negative. Additionally, we have examined the wave crest distribution for various Ursell parameter values, comparing it with the Rayleigh distribution. Our analysis reveals that as the Ursell number increases, the probability distribution function deviates slightly from the theoretical Rayleigh distribution, indicating differences in the distributions.


Understanding the behavior and characteristics of ocean internal waves holds great importance for various scientific disciplines, including oceanography, geophysics, and marine ecology. Researchers and scientists delve into the intricate dynamics of these waves to unravel their impact on marine ecosystems and climate patterns. By enhancing our knowledge of internal waves, we can gain valuable insights into the ocean's intricate workings and its role in shaping the Earth's environment.

\section{Acknowledgements}
M.V.F is grateful to IMPA for hosting him as visitor during the 2023 Post-Doctoral Summer Program. E.P. is supported by support by the RNF grant number 19-12-00253

	\section*{Declarations}
	
	\subsection*{Conflict of interest}
	The authors state that there is no conflict of interest. 
	\subsection*{Data availability}
	
	Data sharing is not applicable to this article as all parameters used in the numerical experiments are informed in this paper.

\end{document}